\documentclass[amssymb,12pt]{article}

\usepackage{scicite}
\usepackage{float}

\usepackage{amsmath}
\usepackage{amsfonts}
\usepackage{graphicx}
\usepackage[outdir=./,update,prepend]{epstopdf}
\usepackage{xcolor}

\newenvironment{sciabstract}{%
\begin{quote} \bf}
{\end{quote}}

\newcounter{lastnote}

\title{Cluster formation and self-assembly in stratified fluids: a novel mechanism for particulate aggregation}

\author
{Roberto Camassa$^{1}$, Daniel M. Harris$^{2}$, Robert Hunt$^{1}$,\\ Zeliha Kilic$^{3}$, and Richard M. McLaughlin$^{1}$\\
\\
\normalsize{$^{1}$Department of Mathematics, University of North Carolina, Chapel Hill,}\\
\normalsize{Chapel Hill, North Carolina, 27599, USA}\\
\\
\normalsize{$^{2}$School of Engineering, Brown University,}\\
\normalsize{Providence, RI 02912, USA}\\
\\
\normalsize{$^{3}$Department of Physics and Center for Biological Physics, Arizona State University,}\\
\normalsize{Tempe, Arizona, 85287}\\
}
\date{}

\begin{document} 




\maketitle


\begin{sciabstract}

An extremely broad and important class of phenomena in nature involves the settling and aggregation of matter under gravitation in fluid systems.  Some examples include: sedimenting ``marine snow" particles in lakes and oceans (central to carbon sequestration) \cite{MacIntyre}, dense microplastics in the oceans (which impact ocean ecology and the food chain \cite{plastic}), and even ``iron snow" on Mercury \cite{mercury} (conjectured as its magnetic field source).  These fluid systems all have stable density stratification, which is known to trap particulates \cite{MacIntyre,Abaid,Camassa1,Camassa2,POFLetter2} through upper lightweight fluid coating the sinking particles, thus providing transient buoyancy.  The current understanding of aggregation of such trapped matter involves collisions (due to Brownian motion, shear, and differential settling) and adhesion \cite{aggregate}.  Here, we observe and rationalize a new fundamental effective attractive mechanism by which particles suspended within stratification may self-assemble and form large aggregates without need for short range binding effects (adhesion).  This phenomenon arises through a complex interplay involving solute diffusion, impermeable boundaries, and the geometry of the aggregate, which produces toroidal flows.  We show that these toroidal flows yield attractive horizontal forces between particles.  We observe that many particles demonstrate a collective motion revealing a system which self-assembles, appearing to solve jigsaw-like puzzles on its way to organizing into a large scale disc-like shape, with the effective force increasing as the collective disc radius grows.  Control experiments with two objects (spheres and oblate spheroids) isolate the individual dynamics, which are quantitatively predicted through numerical integration of the underlying equations of motion.  This new mechanism may be an important process in formation of marine snow aggregates and distribution of phytoplankton in lakes and oceans \cite{phyto}.  Further, it potentially provides a new mechanism for general sorting and packing of layered material.

%

\end{sciabstract}

\vspace{.2 in}
\noindent
{\large \bf Main}
\vspace{.2in}

\begin{figure}[!ht]
\centering
\includegraphics[width=1\linewidth]{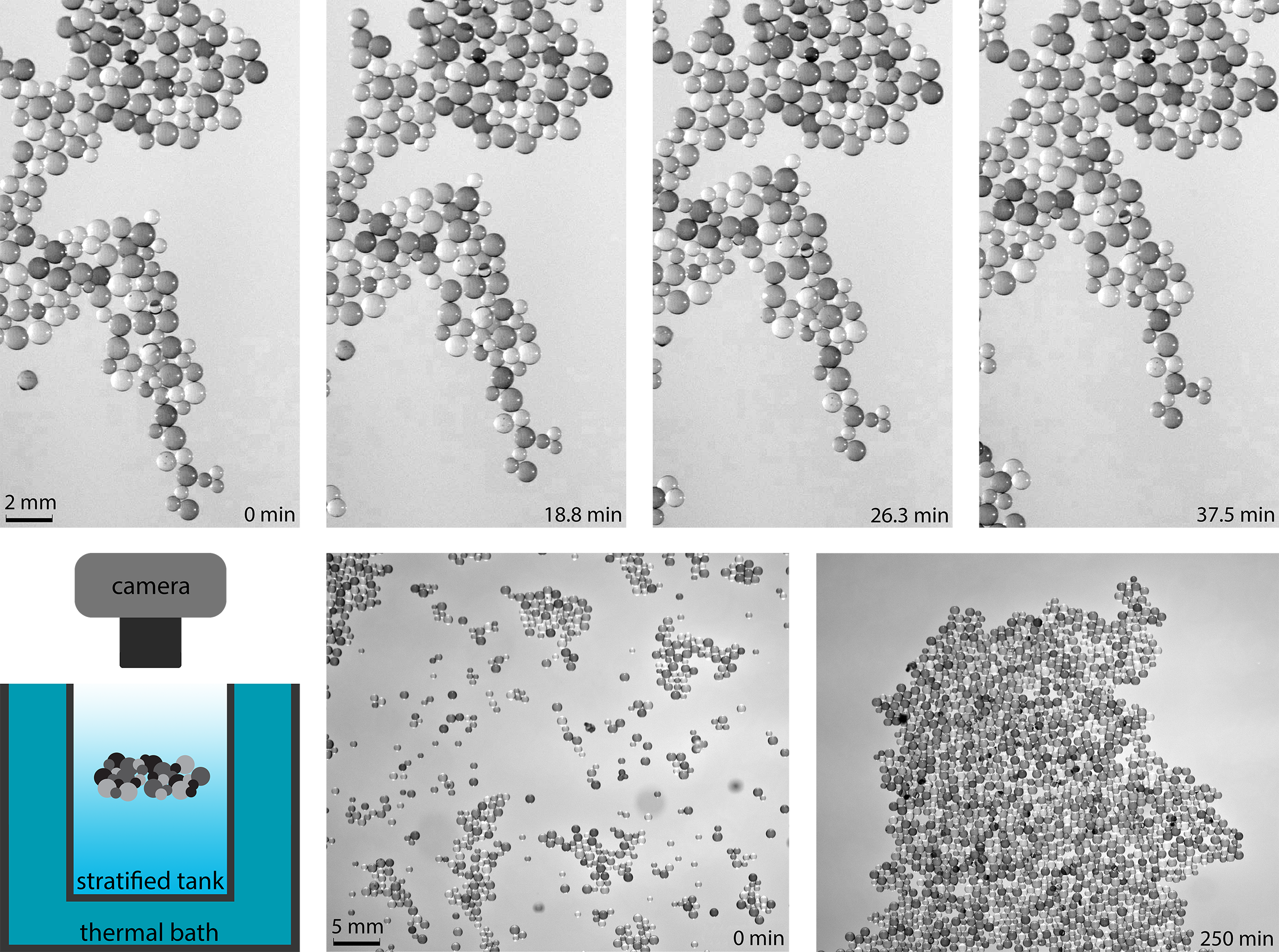}
\caption{Top: Time series of self assembly of collection of neutrally buoyant spheres suspended within a sharply salt stratified fluid viewed from above.  Spheres radii and densities are $0.025-0.05$cm and $1.05$g/cc, top fluid is fresh water ($0.997$g/cc), bottom is NaCl water solution of density ($1.1$g/cc).  Bottom left: Schematic showing experimental setup, bottom center, initial cluster (different trial), bottom right: final cluster.}
\label{montage}
\end{figure}

The self-assembly phenomenon is clearly observed in our experiments (Figure \ref{montage}) with a collection of neutrally buoyant, small spheres suspended at the same height between layers of sharply salt-stratified water in a rectangular plexiglass container.  The spheres are photographed from above (see schematic in Figure \ref{montage}) at regular time intervals.  The spheres, initially isolated, feel a mutually attractive force which forms local clusters.  In turn, these clusters attract each other while orienting themselves to seemingly try to fill a puzzle-like pattern, ultimately resulting in a large disc-like shape.  See Supplementary Video 1 for a dynamic view of this self-assembling cluster.

\begin{figure}[!ht]
\centering
\includegraphics[width=.37\linewidth]{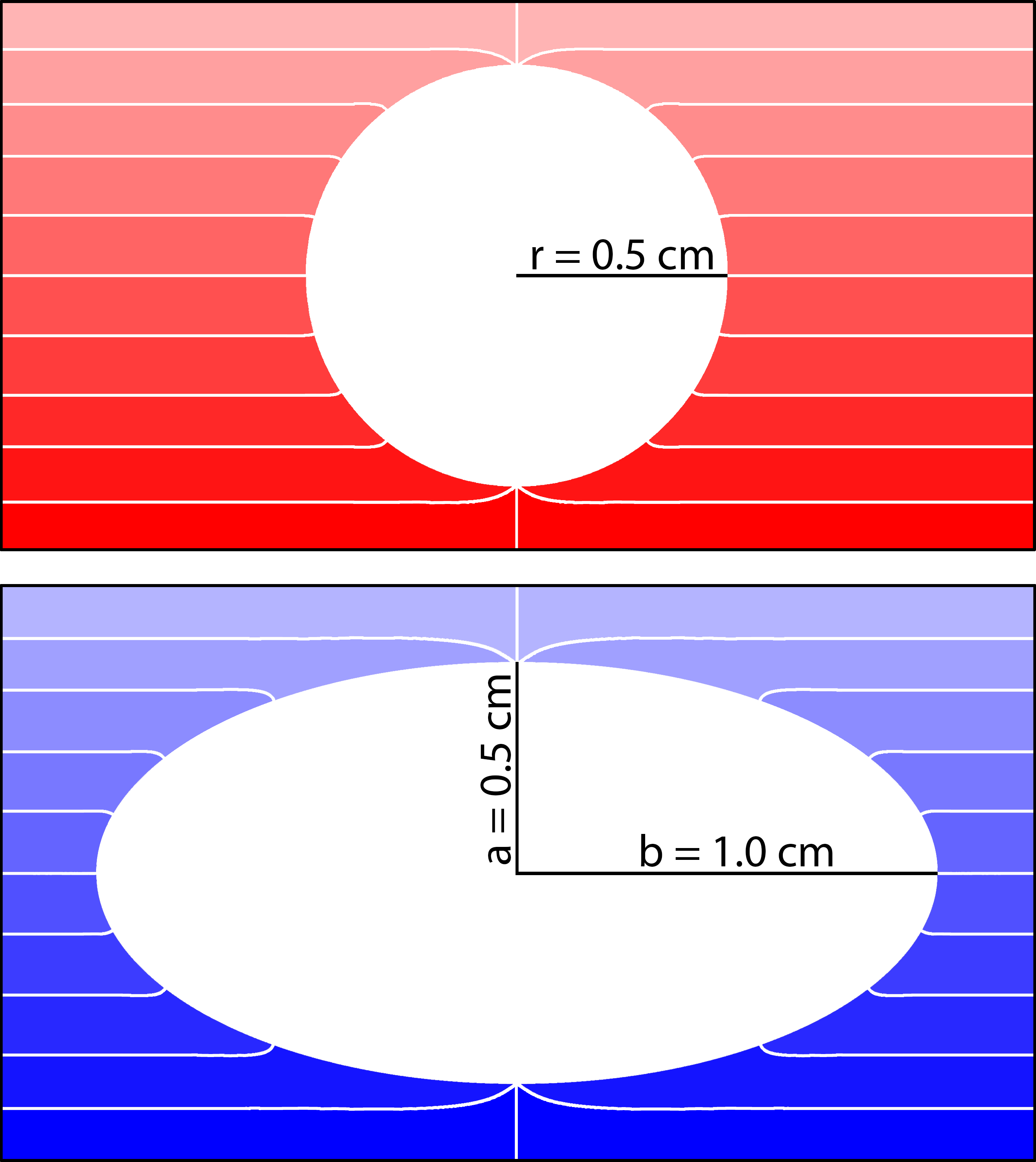}
\includegraphics[width=.6\linewidth]{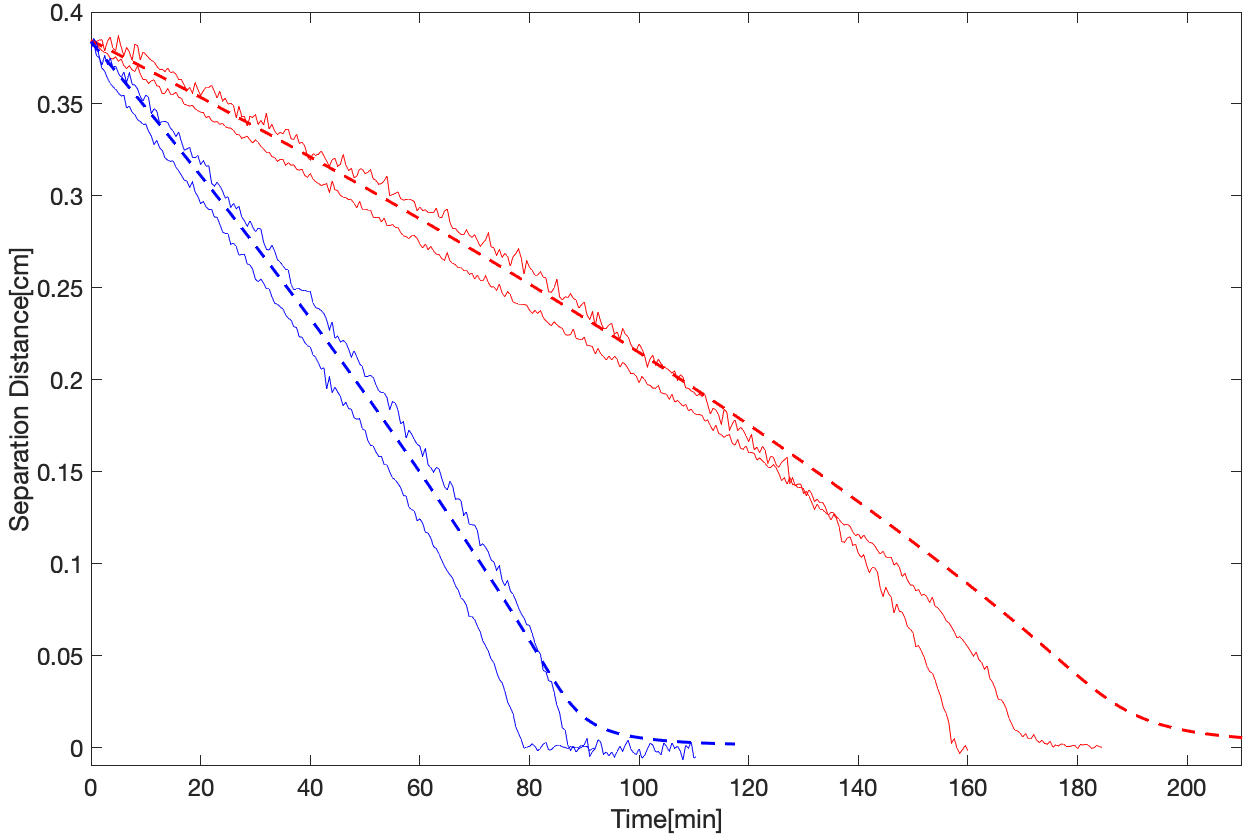}
\caption{Experiments and computations with large sphere/spheroid:  Left panel: side view geometries, and computationally computed density isolines, $Pe=41$, sphere radius, $0.5$cm, spheroid width, $2$cm, density increments approximately $0.001$g/cc, with density gradient $\sigma=0.007$g/cc/cm.  Right panel: Experimental and computational evolution of separation distance as a function of time between a large sphere and a small passive sphere and between a large spheroid and a small passive sphere.  All experimental trajectories are plotted with the same initial separation. Also shown are the companion trackings produced by our computational simulation for the two geometries (dashed lines).  The diffusion induced flows generated by either large body advect and attract the small passive particle until contact.  Note that the velocities in the case of the oblate spheroid are larger than the case of the sphere.}
\label{two}
\end{figure}

To probe more directly the nature of the attractive force between spheres, we next examine cases involving single large bodies attracting single small bodies in linear stratification, which can be assumed to be advected by the fluid flow induced by each individual large body.  All bodies have the same approximate densities ($1.05$g/cc) and float at the same height within the salt stratified layer (density gradient approximately $0.007$g/cc/cm).  The left panel of Figure \ref{two} shows the side view of the two different large bodies considered.  The sphere has radius $0.5$cm.  The oblate spheroid has same vertical radius, while its horizontal radius is $1.0$cm.  We monitor the separation distance as a function of time between the large body and the small body.  In the right panel of Figure \ref{two}, we present these trackings for the case of a large sphere and for the case of a large oblate (horizontal disc-like) spheroid for multiple trials of each experiment.  The small sphere in both cases has a radius of $0.05$cm.  All cases indicate an attractive force is created by the large body upon the small body.  Observe that the slopes of these curves indicate that the attraction is stronger for the wide spheroid than that of the sphere.  This supports the fact that larger discs induce larger attractive forces on smaller particles and gives insight into the collective dynamics of the many particle system:  as clusters grow into large scale discs, their attractive force upon individual particles (or smaller clusters) becomes effectively stronger and stronger until all particles are packed into the large disc-like shape observed in Figure \ref{montage}.  See Supplementary Video 2 for video documenting the collapse for these two cases involving the large sphere/spheroid.

The theoretical explanation for this novel effective force of attraction lies within diffusion induced flows.  These are flows which originate from a mismatch between the background horizontal isolines of density which initially intersect the surface non-orthogonally. The no-flux boundary condition for salt transport implies that this angle of intersection {\bf \em must} be 90 degrees.  As a result, on an initial transient timescale the isolines are bent to locally align with the surface normal.  But this produces a lifting (or depression) of density which creates a buoyancy imbalance which, in turn, creates a fluid flow.  Of course, the full, quantitative explanation of this behavior relies upon the integration of the partial differential equations (PDEs) underlying the fluid system.  While our studies involve fully three-dimensional flows and non-planar geometry, some intuition can be found for the special case of an infinite tilted flat plane inserted into a linearly stratified fluid: for this geometry an exact steady solution to the Navier-Stokes equations coupled to an advection-diffusion equation for salt concentration was derived by O. M. Phillips \cite{Phillips} and C. Wunch \cite{Wunch}.  This solution shows a boundary layer region in which there is a density anomaly and a shear flow up the top side of the sloped wall.  We remark that experimental work of Olhaus and Peacock, using a freely suspended wedge-shaped object, demonstrated that such flows are in fact sufficient to self transport a single object\cite{Peacock}.  However, interactions between multiple objects through diffusion induced flows have not been explored previously.
For our studies, the sphere and spheroids are symmetric, and thus no self-induced motion is generated by a single body in isolation.  And yet, self-induced flows are generated by these bodies, and those flows induce the collective motion of other nearby bodies, such as documented in Figure \ref{montage}.  To study these flows in detail, we will consider first the case of a single body held fixed within a background linear stratification.  

\begin{figure}[!ht]
\centering
\includegraphics[width=1\linewidth]{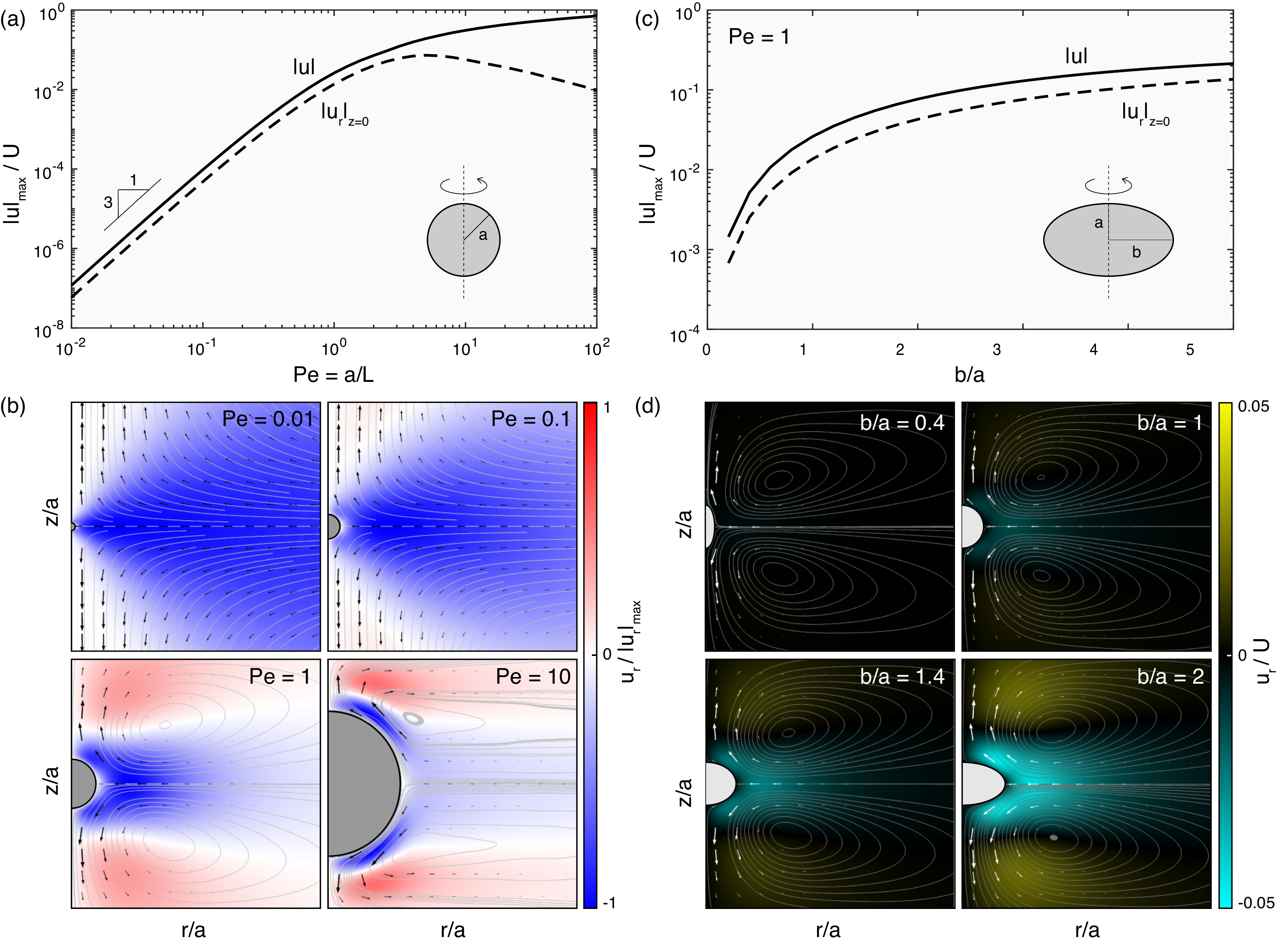}
\caption{Flow strengths and structures for spheres (left panels) and spheroids (right panels), a) Overall max velocity and equatorial maximum cylindrical radial speed, $u_r$ as a function of the Peclet number, b) flow structures induced by a sphere for four different Peclet numbers in a vertical plane slicing the sphere through north and south pole, color representing horizontal velocity scaled by its maximum value for each Peclet number, c) Overall max velocity and equatorial maximum cylindrical radial speed as a function of the aspect ratio, d) flow structures induced by four different spheroids, scaled by the Phillips velocity, $U$.}
\label{flows}
\end{figure}

The characteristic velocity and length scales for the Phillip's solution are $U = \kappa \left( \frac{g
\sigma}{\kappa \mu} \right)^{1 / 4}$, and $L = \left( \frac{\kappa
\mu}{g \sigma} \right)^{1 / 4}$, where $g$ is that gravitational acceleration, $\mu$ and $\kappa$ are the dynamic fluid viscosity and salt diffusivity, and $\sigma$ is the slope of the background density field (all assumed to be constant in this work).  We use this velocity scale to nondimensionalize the equations of motion for the velocity field ${\textbf u}$, pressure $P$, density $\rho$ (varying only through the evolution of the salt concentration), position ${\textbf x}$, and time $t$, (for a single sphere of radius $a$ for brevity in exposition) via $\tilde{\textbf x} =\frac{{\textbf x}}{a}, \tilde{\textbf u}  =  \frac{{\textbf u}}{U}, \tilde{\rho} =\frac{\rho}{\sigma a}, \tilde{P}  = \frac{P a}{\mu U}$, and $\tilde{t}  =  \frac{t \kappa}{a^2}$.
The resulting non-dimensional PDE system (dropping tildes and primes) for the incompressible fluid velocity and concentration field is:
\begin{eqnarray}
  Re \ \rho \left[ \frac{1}{Pe} \frac{\partial {\textbf u}}{\partial t} +
  {\textbf u} \cdot \nabla {\textbf u} \right] & = & - \nabla P - 
  Pe^3 \rho \hat{z} + \Delta {\textbf u}
  \label{momentum}\\
  \frac{\partial \rho}{\partial t} + Pe \ {\textbf
  u} \cdot \nabla \rho  & = & \Delta \rho
\end{eqnarray}
where the Reynolds number is $Re =  \frac{\sigma a^2 \bar{u}}{\mu}$ and the Peclet number is 
$Pe =  \frac{a}{L}$.  For all the experiments we have run, $Re<0.001$, and, as such, the so-called Stokes approximation may be employed which sets the left hand side of equation (\ref{momentum}) to zero while retaining nonlinear effects through non-zero $Pe$ in the advection-diffusion of the salt concentration.  As such, the only remaining parameter in the system is the Peclet number.  The boundary conditions are assumed to be no-slip for the velocity, and no-flux for the tracer on any physical boundary.  

We seek steady solutions of these equations using two methods.  The steady approximation is well supported through standard partial differential equation estimates, which 
suggest that the transients will decay after the first hour of our experiment (typically lasting $16$ hours). First, for general Peclet numbers, we utilize the finite element package COMSOL to calculate steady solutions of these equations of motion.  See the Supplementary Materials section for extended details on this calculation.  Second, for very small Peclet numbers, this evolution of the salt concentration decouples from the fluid velocity and in fact is, surprisingly, equivalent to finding the irrotational velocity potential for the velocity field induced by a steady vertical flow past a fixed sphere or spheroid.  In turn, the actual velocity field is constructed through convolution of the concentration field, $\rho$, with the available Green's function associated with a point source of momentum in the presence of a no-slip sphere.  See the Supplementary Materials section for details regarding these solutions, the Green's function, and the imaging methods employed to generate physical solutions.  

Figure \ref{flows} depicts the simulations for the former method, documenting the toroidal flow structures observed for a range of Peclet numbers.  In the top left panel, we document, for the case of a sphere, the maximum flow speed in the equatorial plane as well as the maximum speed overall in the entire domain as a function of the Peclet number.  
In the equatorial plane, there is an optimal velocity which occurs at a Peclet number slightly bigger than unity.  As a benchmark, observe that the slope (here in log-log coordinates) at low Peclet number is $3$ (i.e. ${\bf u}\sim Pe^3$), which is the exact theoretical scaling in this limit.  In the lower left panels, the companion flow structures are observed for four representative Peclet numbers.  The background color scheme is assigned by the radial velocity, with negative values corresponding to horizontal velocities directed left.  Observe that in all cases, in the equatorial plane, the velocities are directed to attract nearby fluid particles towards the sphere.  In the right panels of Figure \ref{flows} we show the analogous plots for spheroids of different aspect ratios.  In the top, we show the max speed in the equatorial plane and global max speed as a function of aspect ratio, while in the bottom, we document the toroidal flow structures for four representative aspect ratios.  In all cases, we observe a monotonic increase in the max equatorial speed as the horizontal disc radius increases.  We next apply this method to the experimental observations reported in Figure \ref{two}.   For this case involving NaCl in water, $\kappa \simeq 1.5*10^{-5}$cm$^2$/s, the radius $a=0.5$cm, and $Pe\simeq 41$.  By assuming the smaller sphere moves as a passive tracer under the flow induced by the large sphere  (or spheroid), we compute the evolution of the separation distance between the large and small bodies.  These predictions are shown superimposed over the experimental data in the right panel of Figure \ref{two}, and we also show the corresponding computed density isolines in the left panel for the sphere/spheroid.  The agreement is quantitative until the objects get sufficiently close together, at which point the passive tracer assumption begins to fail as particle-particle interactions develop.

\begin{figure}[!ht]
\centering
\includegraphics[width=1\linewidth]{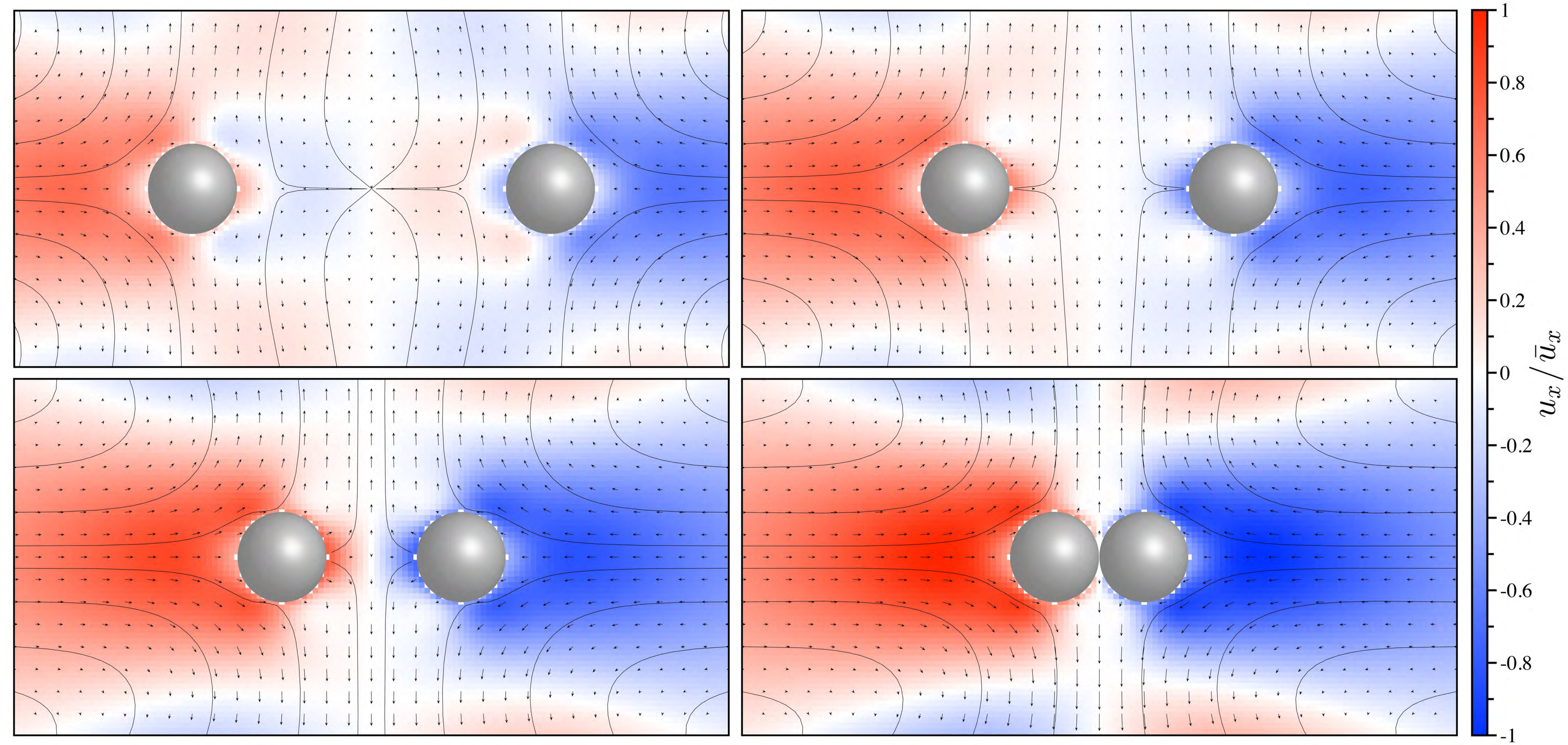}
\caption{Flow structures created by two spheres, at four different fixed separation distances ($6,4,2$, and $0$ radii), with flows computed in the low Peclet number asymptotic limit, scale bar on the right normalized by the maximum of max horizontal speed over the four separation distances.}
\label{lowPe}
\end{figure}

To explore these particle-particle interactions, we work within the analytically tractable low Peclet limit using the method of images.  See Supplementary Material for extended details regarding the Green's function and how the imaging is performed.  Figure \ref{lowPe} displays the flows induced by two fixed spheres in the low Peclet limit, for four different separation distances.  The trend is that for well separated spheres, interactive effects are minimal, as the flow structures resemble the case for an individual from Figure \ref{flows}.  As the spheres are moved closer together, the sphere-sphere interaction modifies the flow structures, with the flow immediately between the spheres weakening as they approach each other (note the method of images induces some mild errors near the boundaries, but is faithful away from boundaries).  Ultimately, as the spheres touch, the flows qualitatively resembles that of a single prolate spheroid, and produces the strongest flows over the four cases.  

In this letter, we have demonstrated a new force of attraction which exists between bodies suspended in a stratified fluid.  
The force origin lies in a body's self creation of a diffusion induced, toroidal flow which attracts nearby matter suspended at the same depth.  Our analysis has shown that these flows are quantitatively predicted by solving the Stokes equations coupled to an advection-diffusion equation.  We remark that we have also experimentally observed that these flows may induce an effective {\it repulsive} force for particles suspended above (or below) a large body, as they are on the opposite (repulsive) side of toroid (see Supplementary Video 3).  We also stress that this new phenomena occurs over a wide range of different particle materials (glass, photopolymer resin, polystyrene), and we have even observed the same behavior in a non-electrolytic fluid (watery corn syrup).  The next steps will be directed towards full prediction of the self-assembly, which involves computing the full fluid-structure interactions between the bodies and the fluid.  Having the first principle theory quantitatively predicting the source of this attractive force provides an advantage in better understanding self-assembly processes over other systems involving active matter whose predictive theories are less certain.  We remark that our new mechanism is distinct from other small scale mechanisms (electro/diffusiophoresis) in that the particle motion is {\em orthogonal to the background gradient} (with velocities scaling sublinearly in $\sigma $ as opposed to linearly \cite{phoresis}).  Further, for the case of symmetric bodies, body motion is only induced through collective interactions.  We also emphasize that this process is expected to be relevant in oceans and lakes, particularly at scales smaller than the Kolmogorov scale of turbulence.  Below this scale, which can vary from centimeters (at depth) to tens of microns (in surface water) \cite{thorpe,aggregate}, turbulence is negligible, and these processes will participate in the formation and clustering of marine aggregates.  
 
\vspace{.2 in} 
\noindent
\textbf{Methods}
\vspace{.2in}

To create a linear stratification, deionized water is mixed with NaCl, degassed in a vacuum chamber at -24inHg for 10 minutes, and thermalized in a water bath. A two bucket method is used, where fresh water is pumped via Cole-Parmer Ismatec ISM405A pumps at a constant rate into a mixing bucket initially full of saline. Concurrently, this mixture is pumped through a floating diffuser to a depth of approximately $10$cm inside a tank $30.5$cm tall, $16.5$cm wide, and $15.5$cm deep submerged in a custom built Fluke thermal bath to regulate temperature. The short depth walls are made of $1/4$ inch thick copper to facilitate thermal coupling, and the longer width walls are made of $7/16$ inch acrylic to allow for visualization.  To measure the density profile, a Thermo Scientific Orion Star A215 conductivity meter equipped with an Orion 013005MD conductivity cell is carefully submerged into the tank along the wall using a linear stage that measures with $0.01$mm precision. 

Density matched objects (spheres and/or spheroids of approximately $1.05$ g/cc with radii varying from $50$ microns to centimeter scale)  are lowered into the fluid slowly using a thin wire support, with care taken to remove any air bubbles on the surface of the body. Once they have been placed at their buoyant height near the center of the tank, the stratified tank is left for a minimum of $12$ hours to allow decay of transients.  From above, images are acquired at $15$, $30$, or $60$ second intervals using a Nikon D3 camera equipped with an AFS Micro Nikkor 105mm lens and a Nikon MC-36A intervalometer. The tank is covered on top with a $1/4$ inch acrylic sheet to reduce evaporation and prevent convective motion. 

To manufacture spheres and spheroids, we used a Formlabs Form 2 3D printer with Formlabs Clear V4 photopolymer resin printed at $25$ micron layer thickness. The bodies were printed in halves split along the equatorial plane and joined together with epoxy. An interior cavity in the top half allows for the printed bodies to be density matched to polystyrene tracer spheres of density $\simeq 1.05$ g/cc.  The polystyrene spheres used for the self-assembly experiments varied in radius from $0.025$cm to $0.05$cm, and the vertical density profile was two-layer (rather than linear as used in the control experiments).  Moreover, we note that two sphere experiments were performed with glass spheres of radius $0.25$ cm and similar attraction was observed.  We also note that similar attractions were observed in experiments performed in a non-electrolytic corn syrup solution with density stratification achieved by varying water content.

\textbf{Acknowledgments} We acknowledge funding received from the following NSF Grant Nos.: 
RTG DMS-0943851, CMG ARC-1025523, DMS-1009750, and DMS-1517879; and ONR Grant No: ONR N00014-12-1-0749.

\bibliographystyle{Science}

\noindent


%

\end{document}